\documentclass[prl,aps,twocolumn,superscriptaddress]{revtex4-1}
\usepackage{amssymb,amsfonts,amsmath,amstext,graphicx,hyperref}

\hypersetup{
colorlinks=true,
linkcolor=blue,
filecolor=magenta,  
urlcolor=cyan,
}
\usepackage[dvipsnames]{xcolor}

\begin{document}

\title{Comment on ``Noise, not squeezing, boosts synchronization in the deep quantum regime'' arXiv:2002.07488 (2020)}

\author{Michal Hajdu\v{s}ek}
\email{michal@keio.jp}
\affiliation{Keio University Shonan Fujisawa Campus, 5322 Endo, Fujisawa, Kanagawa 252-0882, Japan.} 

\author{Sai Vinjanampathy}
\email{saiv@phy.iitb.ac.in}
\affiliation{Department of Physics, Indian Institute of Technology-Bombay, Powai, Mumbai 400076, India.}
\affiliation{Centre for Quantum Technologies, National University of Singapore, 3 Science Drive 2, 117543 Singapore, Singapore.}

\date{\today}

\maketitle

\makeatletter

In reference \cite{mok2020noise}, the authors consider a driven quantum van der Pol oscillator described by
\begin{equation}
\dot{\rho} = -i [H, \rho] + \gamma_1 \mathcal{D}[a^{\dagger}] \rho + \gamma_2 \mathcal{D}[a^2] \rho + \kappa \mathcal{D}[a] \rho,
\label{eq:master_equation}
\end{equation}
where $H = \delta a^{\dagger}a + \Omega (a + a^{\dagger}) + \eta (a^2 + a^{\dagger 2})$.
Similar system was considered in \cite{sonar2018squeezing}, the difference being that reference \cite{mok2020noise} includes a single-photon relaxation at rate $\kappa$.
The authors of reference \cite{mok2020noise} consider the behaviour of master equation (\ref{eq:master_equation}) in the deep quantum limit when $\gamma_2/\gamma_1\rightarrow\infty$.
The two major claims in reference \cite{mok2020noise} are that in this limit (i) squeezing is not a good tool to entrain the oscillator, and (ii) that single-photon relaxation at rate $\kappa$ boosts synchronization in this limit.

In this comment we show that Eq.~(\ref{eq:master_equation}) is not a correct description of the system in the deep quantum limit. We also address both of the major claims and show that they are not novel.
We then point out serious inconsistencies and misleading statements in the authors' arguments.

\section{Consistency of Master equations}
Before addressing the main results of \cite{mok2020noise} we wish to discuss the validity of the so-called ``deep quantum limit" and why it is physically dubious.  

A standard method to derive quantum master equations is associated with Davies \cite{davies1,davies2,davies3}, wherein we begin with a system-bath Hamiltonian $H_{SB}=H_S+H_B+\lambda H_{SB}$ and move to the interaction picture.
After that, the derivation proceeds by making several crucial approximations, namely weak coupling approximation, Born, Markov, RWA and the so-called Davies or van Hove limit \cite{davies1,davies2,davies3,dumcke1979proper}.
We focus our attention on the van Hove limit, which demands that the coupling strength $\lambda\rightarrow 0$ and $t\rightarrow \infty$ such that $\lambda^2 t\rightarrow \text{constant}$.
If this so-called van Hove limit is satisfied, then the Markovian master equation is valid and the steady states truly represent the underlying dynamics of the system.
Immediately it becomes clear that $\lambda\rightarrow\infty$ is a disallowed limit for the validity of time local Markovian physics.
Now, typically the bath itself is taken to be in a Gibbs state, and produces the product of two terms.
One of these terms is $\lambda^2=\gamma$ and the other is the Fourier transform of the bath-bath correlation function, which typically produces a temperature dependent term of the form $(n_{Th}+1)$ for the cold bath and $n_{Th}$ for the hot bath.
In master equations like the vdP master equation under consideration in reference \cite{mok2020noise}, there are non-thermal dissipators usually thought to come from two zero temperature baths which are engineered by setting $T\rightarrow 0$ or $\beta\rightarrow\infty$.
This causes each bath to only produce one term (say linear pumping for one bath and non-linear damping for the other bath), which is what is presented as the model. 

This raises an issue which can be understood by considering the Gibbs state of the system at any temperature $\beta^{-1}$, which is given by the partition function $\mathcal{Q}_S=\mathcal{Q_{SB}}/\mathcal{Q_B}$. Here  $\mathcal{Q}_{SB}=tr(e^{-\beta H_{SB}})$ and $\mathcal{Q}_B=tr(e^{-\beta H_{B}})$ \cite{campisi2009fluctuation}.
If the limit $\beta\rightarrow \infty$ is taken, then all levels of perturbation have to be taken into account to get correct equilibrium physics.
Another way of saying this is that at zero temperature, the validity of a second order perturbation theory has to be carefully considered.
Typically this is not given serious consideration since bath and system parameters are assumed to be within the weak coupling and van Hove limit. 

This brings us to the main concern about the validity of the ``deep quantum regime".
If you take $(\lambda_2/\lambda_1)^2\rightarrow\infty$ the van Hove limit is inconsistent. Furthermore, if simultaneously $\beta\rightarrow\infty$, then the validity of the perturbative expansion that leads to the Lindblad master equation has to be inspected, an issue that is further complicated by several baths. To avoid all these inconsistencies in relation to deriving master equations, authors previously stayed away from large values of $\gamma_2/\gamma_1$. The authors' \cite{mok2020noise} claims should hence be considered carefully for real physical systems, since if $\gamma_2/\gamma_1\rightarrow\infty$, the master equation most certainly will differ from simple Lindblad type.

\section{Specific Claims}
\textit{Claim (i).---}
The claim of reference \cite{mok2020noise} that squeezing does not work in the authors' definition of deep quantum limit is not new.
This is clear from the form of the steady state of Eq.~(\ref{eq:master_equation}) of the quantum van der Pol oscillator, namely $\rho_{ss}=\frac{2}{3}|0\rangle\langle0|+\frac{1}{3}|1\rangle\langle1|$, which has been reported in \cite{lee2013quantum}.
Reference \cite{sonar2018squeezing} studies synchronization when $\gamma_2/\gamma_1=3$ for precisely this reason.
In this regime the steady state in \cite{sonar2018squeezing} has less than one photon on average and is characterised by sub-Poissonian statistics but still has considerable populations beyond the lowest two levels allowing squeezing to be effective.
Reference \cite{mok2020noise} tries to extend the analysis presented in \cite{sonar2018squeezing} to the limit of $\gamma_2/\gamma_1\gg 10$ where the validity of the master equation (\ref{eq:master_equation}) has to be first established. Ignoring this issue about the validity, the claim that squeezing is ineffective in \textbf{this} regime does not imply that it is ineffective in the original $\gamma_2/\gamma_1\approx3$ regime considered. \textbf{This subtle point is made on page 4 by the authors, though the title of the paper implies otherwise, which is scientifically misleading and a serious misrepresentation of the results of both papers.}

\textit{Claim (ii).---}
The other major claim of reference \cite{mok2020noise} is that noise boosts synchronization.
Again, this claim is not particularly novel.
Reference \cite{roulet2018synchronizing} considers an externally driven spin-1 system and explicitly shows that synchronization measure is $S(\phi)\approx(1/\gamma_g-1/\gamma_d)\cos(\phi)$, with $\gamma_g$ and $\gamma_d$ being the incoherent rates of energy gain and dissipation, respectively.
One can immediately see that $\gamma_g=\gamma_d$ leads to no synchronization and adding noise by increasing $\gamma_d$ results in increase of synchronization.
Similar effect has been also observed in \cite{jaseem2020quantum} in the context of synchronization in nanoscale heat engines.
Here a general three-level atom may be synchronized to an external drive when operated as an engine.
Adding noise by increasing the temperature of one of the baths decreases synchronization which vanishes at the Carnot point.
Adding further noise pushes the heat engine into the fridge regime and results in a finite synchronization again.
Therefore both major results of reference \cite{mok2020noise} are a recycling of recent but well established results in the field of quantum synchronization.

\begin{figure}
    \centering
    \includegraphics[width=\columnwidth]{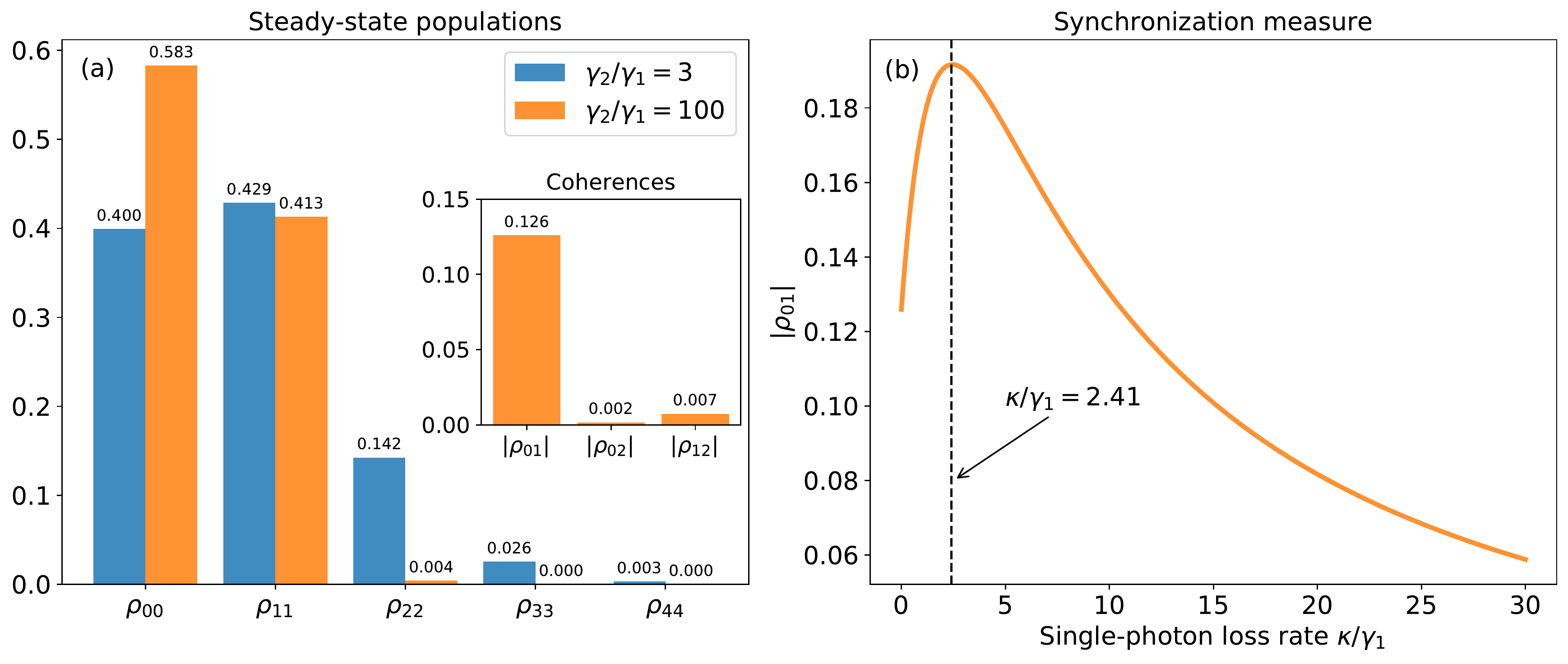}
    \caption{(a) Comparison of steady-state populations of an undriven van der Pol oscillator. Blue bars represent the case of $\gamma_2/\gamma_1=3$ considered in reference \cite{sonar2018squeezing}. Orange bars represent the case of $\gamma_2/\gamma_1=100$ investigated in reference \cite{mok2020noise}. The magnitudes of coherences when $\gamma_2/\gamma_1=100$ are shown in the inset. Other parameters are $\delta/\gamma_1=0.1$, $\kappa/\gamma_1=0$, $\Omega/\gamma_1=1$, and $\eta/\gamma_1=0$. (b) Synchronization measure of reference \cite{mok2020noise} as a function of increasing single-photon loss rate $\kappa/\gamma_1$. The synchronization is initially boosted but then decreases. Other parameters are $\Omega/\gamma_1=1$, $\delta/\gamma_1=0.1$, $\gamma_2/\gamma_1=100$, $\eta/\gamma_1=0$.}
    \label{fig:figure}
\end{figure}

\textit{Ansatz inconsistency.---}
Authors of reference \cite{mok2020noise} claim that the steady state is well described by Eq.~(3) in \cite{mok2020noise} where they truncate the infinite-dimensional Hilbert space to the lowest three levels and neglect all coherences involving level $|2\rangle$.
Fig.~\ref{fig:figure}(a) shows that the neglected coherences have magnitude of the same order as $\rho_{22}$.
In fact $|\rho_{02}|$ is nearly twice the size of $\rho_{22}$ clearly showing that the ansatz is inconsistent.
Including these coherences would in fact increase any reasonable measure of synchronisation though the analysis needs to be redone carefully since the claimed effect might not persist. We comment on this now.

\textit{Physics behind noise-boosted synchronization.---}
Authors of \cite{mok2020noise} make an attempt at explaining why the incoherent one-photon loss at rate $\kappa$ leads to a boost to synchronization.
They argue that increasing $\kappa$ leads to an increase in the ground state population which in turn produces larger magnitude of coherence $\rho_{01}$.
It is true that larger $\kappa$ leads to an increase in the ground-state population.
Therefore if the reasoning of \cite{mok2020noise} is correct one would expect $|\rho_{01}|$ to be a monotonously increasing function of $\kappa$.
Fig.~\ref{fig:figure}(b) shows that this is not the case and in fact the coherence does decrease in magnitude with increasing $\kappa$ (and increasing population $\rho_{11}$) as one would expect.
The authors present this figure in Fig.~4 of \cite{mok2020noise}, they fail to discuss the fact that noise boosts synchronization for only a very limited range of $\kappa$.
The physics in their deep quantum limit (where the master equation is suspect) is perhaps that while bringing down population from the first excited level to the ground level works for a little bit, the deleterious effect of a linear damp on the populations overwhelms any advantage provided to the coherences.

\textit{Limit cycle size.---}
The authors correctly note that the addition of single-photon loss at rate $\kappa$ decreases the average population in the steady state.
In order to compensate for this loss the authors increase the harmonic driving strength as given by Eq.~(10) in \cite{mok2020noise},
\begin{equation}
\Omega_{th}^2 = \frac{\epsilon (3\gamma_1 + \kappa) (6\gamma_1\kappa + 9\gamma_1^2 + 4\delta^2 + \kappa^2)}{4[ \gamma_1(1 - 6\epsilon) + \kappa(1 - 2\epsilon)]},
\end{equation}
where $\epsilon$ is a free parameter.
The authors of \cite{mok2020noise} set the new driving strength $\Omega=\Omega_{th}$ which increases with the single-photon loss rate $\kappa$.
The authors then observe increase in their synchronization measure with increasing $\kappa$ and conclude that noise is boosting synchronization.
\textbf{This is not the correct way of studying the sole effect of noise on synchronization.}
It is the increasing harmonic drive which results in stronger entrainment of the oscillator yet the authors conclude that this is due to the increasing single-photon rate $\kappa$.
This is the \textit{wrong} conclusion.

\end{document}